\documentclass[12pt,preprint]{aastex}
\usepackage{amsmath}

\shorttitle{Rotation curve \& DM profile of MW}
\shortauthors{Ablimit at el.}

\begin{document}


\title{The rotation curve, mass distribution and dark matter content of the Milky Way from Classical Cepheids}
\author{Iminhaji Ablimit\altaffilmark{1,2,\star}, Gang Zhao\altaffilmark{1},  Chris Flynn\altaffilmark{3},
and Sarah A. Bird\altaffilmark{1}
}
\altaffiltext{1}{Key Laboratory for Optical Astronomy, National Astronomical Observatories, Chinese Academy of Sciences, Beijing 100012,
China. iminhaji@nao.cas.cn; gzhao@nao.cas.cn}
\altaffiltext{2}{Department of Astronomy, Kyoto University, Kitashirakawa-Oiwake-cho, Sakyo-ku, Kyoto 606-8502, Japan. iminhaji@kusastro.kyoto-u.ac.jp}
\altaffiltext{3}{Centre for Astrophysics and Supercomputing, Swinburne University of Technology, P.O. Box 218, Hawthorn, VIC 3122, Australia. cflynn@swin.edu.au}
\altaffiltext{0}{$^\star$LAMOST Fellow}
\altaffiltext{0}{For correspondence, it should be addressed to I. Ablimit and G. Zhao}


\begin{abstract}

With the increasing numbers of large stellar survey projects, the quality and quantity of excellent
tracers to study the Milky Way is rapidly growing, one of which is the classical Cepheids. Classical Cepheids
are high precision standard candles with very low typical uncertainties ($<$ 3\%) available via the mid-infrared period-luminosity relation.
About 3500 classical Cepheids identified
from OGLE, ASAS-SN, Gaia, WISE and ZTF survey data have been analyzed in this work, and their spatial distributions show a clear signature of Galactic warp.
Two kinematical methods are adopted to measure the Galactic rotation curve in the
Galactocentric distance range of $4\lesssim R_{\rm GC} \lesssim 19$ kpc.
Gently declining rotation curves are derived
by both the proper motion (PM) method and 3-dimensional velocity vector (3DV) method. The largest sample of classical
Cepheids with most accurate 6D phase-space coordinates available to date are modeled in the 3DV method, and the resulting rotation
curve is found to decline at the relatively smaller gradient of ($-1.33\pm0.1$) ${\rm km\,s^{-1}\,kpc^{-1}}$. Comparing to results from the PM method, a higher rotation velocity
(($232.5\pm0.83$) ${\rm km\,s^{-1}}$) is derived at the position of Sun in the 3DV method. The virial mass and
local dark matter density are estimated from the 3DV method which is the more reliable method, $M_{\rm vir} = (0.822\pm0.052)\times 10^{12}\,M_\odot$
and $\rho_{\rm DM,\odot} = 0.33\pm0.03$ {GeV\footnote{Units of GeV ${\rm cm^{-3}}$ may be more seen in the particale physics;
For astronomers, there is a useful conversion: 0.008$M_\odot\,{\rm pc^{-3}} = 0.3 \rm GeV {\rm cm^{-3}}$.} ${\rm cm^{-3}}$}, respectively.

\end{abstract}


\section{Introduction}

The mass distribution and dark matter density profiles of the Milky Way are not just key probes of its assembly history
(e.g., Lake 1989; Read et al. 2008; Deason,  Belokurov \& Sanders 2019),
 but also provide crucial clues
for the cosmological context of galaxy formation (e.g., Dubinski 1994; Ibata et al. 2001; Lux et al. 2012).
The two distributions are usually studied in the frame work of the `standard' Cold Dark Matter model (${\Lambda}$CDM for short,
where the ${\Lambda}$ refers to the density of `dark energy'). In this cosmological model, the energy density of
the Universe comprises approximately 5\% of baryons, 27\% of dark matter and 68\% of dark energy.
The rotation (or circular velocity) curve measurement is a classical way to deliver an indirect measurement of
 these profiles of the Milky Way (Volders 1959; Freeman 1970; Bosma \& van der Kruit 1979;
van Albada et al. 1985; Sofue et al. 2009).

Specifically, the Galactic rotation curve (RC) is the mean
circular velocity around the center of the Galaxy as a function of galactocentric distance measured in the disk-mid plane.
The RC is has been derived with various methods and various tracer objects moving in the gravitational potential of the
Galaxy (e.g., Wilkinson \& Evans 1999; Weber \& de Boer 2010; Sofue 2012; Nesti \& Salucci 2013). For example,
the RC of the Galactic inner region has been derived by the tangent-point method associated with $\rm{H}\,{\rm{I}}$
regions (Gunn et al. 1979; Levine et al. 2008; Sofue et al. 2009). Comparing to the tangent-point method,
methods using stars, dwarf galaxies or globular clusters with distances and at least one of the velocity
components (radial velocity and/or proper motions) are considered as the better measurement for the Galactic
(inner and outer regions) RC (e.g., Smith et al. 2007; Honma et al. 2007;
Bovy et al. 2012; Bovy \& Rix 2013; Bhattacharjee et al.2014; Kafle et al 2014; Reid et al. 2014; Bowden et al. 2015;
Binney \& Wong 2017; Pato \& Icco 2017; Russeil et al. 2017; Ablimit \& Zhao 2017; Katz et al. 2018; Monari et al. 2018;
Sohn et al. et al. 2018). Recently, the measured number of tracers with accurate 6 dimensional (D) phase-space information
is increasing rapidly, with the growing numbers of sky surveys, such as SDSS, Gaia, WISE, ZTF, OGLE, ASAS, Gaia-ESO, APOGEE, etc.,
and these data enable us to measure more precise RC.

Certain types of variable stars are excellent distance indicators due to well-known period-luminosity relations.
Thus, they are taken as excellent objects to study the structure, kinematics and dynamics of the Galaxy,
such as RR lyrae stars (Ablimit \& Zhao 2017; Medina et al. 2018; Ablimit \& Zhao 2018; Utkin et al. 2018; Wegg et al. 2018)
and Cepheids (e.g., Kawata et al. 2018). Frink, Fuchs \& Wielen (1995) derived the Galactic rotation curve from proper motions of 144 Cepheids. Subsequently,
 Pont et al. (1997) constructed the RC of the Galaxy from radial
velocities of 48 classical Cepheids distributed in the outer disc region between the Galactocentric distance 10 kpc and 15 kpc. Gnaci$\acute{\rm n}$ski (2019) obtained the RC by
adopting three kinematic approaches by using 160, 228 and 120 classical Cepheids from the catalog of Mel'nik et al.(2015).
They showed that the slope of the RC lies between a flat RC and a Keplerian
rotation curve. However, Mr$\acute{\rm o}$z et al. (2019) tracked the RC from the 6D phase-space
information of 773 classical Cepheids, and they found a relatively flat rotation curve.
They did not estimate mass distribution and dark matter content of the Milky Way.

In this work, we have selected and analyzed about 3500 classical Cepheids which have precise
distances and measured the Milky way rotation curve using the proper motion method (Gnaci$\acute{\rm n}$ski 2019) and 3D velocity
vector method (Reid et al. 2009). In \S 2, we introduce the classical Cepheids data collected for this work.  Two
methods to calculate the rotation velocities of classical Cepheids are introduced, and the resulted
rotation curve \& its constraint on the mass and dark matter profile of our Galaxy are
given and discussed in \S 3. The concluding remarks are presented in \S 4.


\section{Data Selection}
\label{sec:model}

We collected our sample from several classical Cepheid catalogs as follows: the All-Sky Automated Survey for Supernovae (ASAS-SN) Variable stars catalog
(Shappee et al. 2014; Jayasinghe et al. 2018),  the classical Cepheids sample by
Skowron et al. (2019a, b) basically from the Optical Gravitational Lensing Experiment (OGLE) (Udalski et al. 2015; Udalski et al. 2018), classical Cepheids
from the European Space Agency (ESA) mission Gaia (Gaia Collaboration 2016, 2018; Ripepi et al. 2019),
and the classical Cepheids catalog by Chen et al.(2019) from the Wide-field Infrared
Survey Explorer (WISE) (Wright et al. 2010).
We added new
classical Cepheids identified from the Zwicky Transient Facility (ZTF) catalog (Bellm et al. 2019) by Chen et al.(2020).
We made a cross-match of all the Cepheids from different catalogs in order to remove multiple entries.
Then, we selected Cepheids which have mid-infrared ($W1,W2,W3$ and $W4$ bands) magnitudes from the All WISE catalogue. We calculated
heliocentric distances ($D_{\rm h}$) based on the relations given
in Wang et al. (2018) with $W1,W2,W3$ and $W4$ bands, and took average values for each Cepheid (also see Skowron et al. (2019a) for same calculation method).
Recently, it has been discussed that
 distances derived from mid-infrared period-luminosity relations are more accurate than distances obtained
from parallaxes (e.g., Mr$\acute{\rm o}$z et al. 2019).
After deriving distances, we keep classical Cepheids with $|z|\le 4$ kpc, and we have 3483 classical cepheids
(Galactic longitude ($l$) and latitude ($b$) distributions are shown in the upper left panel of Figure 1): 2223 of them
from Skowron et al. (2019a, b)(magenta stars and red circles),
160 from ASAS-SN catalog (blue squares), 303 from Gaia catalog (open Violet left triangles), 167 from Chen et al.(2019) (green triangles),
618 of them are from Chen et al.(2020) (black triangles).

The spatial distributions are shown in the Figure 1, and all distributions show the clear Galactic warp
which is reported by Skowron et al. (2019a, b) and Chen et al.(2019). The 3D positions of Cepheids and
galactocentric distances ($r$) in the Cartesian coordinate system are calculated as
$x = {\rm R}_\odot - D_{\rm h}{\rm cos}\,b\, {\rm cos}\,l$, $y = D_{\rm h}{\rm cos}\,b\, {\rm sin}\,l$,
$z = D_{\rm h}{\rm sin}\,b$ and $r = \sqrt{x^2 + y^2 + z^2}$,
where ${\rm R}_\sun$ is the distance from the Sun to the Galactic center, and the recent most accurate
value, $8.122\pm0.031$ kpc (GRAVITY collaboration et al. 2018), is adopted.
The projection of galactocentric
distance on the Galactic plane ($R$) is as follows,

\begin{equation}
R = \sqrt{x^2 + y^2}.
\end{equation}


\section{Modeling the rotation curve}

\subsection{The halo model}

The rotation velocity at a radius $R$ from the center of an axisymmetric mass distribution
is related to the total gravitational potential within $R$ and mass $M(<R)$ (at $z\sim 0$),

\begin{equation}
V^2_{\rm c} (R) = R\frac{\partial \Phi}{\partial R} = \frac{GM(<R)}{R},
\end{equation}
where $ \Phi$ and $G$ are the gravitational potential and gravitational constant, respectively.
If we consider the bulge, thin disk, thick disk and dark matter halo for the Galactic potential,
which for the respective contributions are ${\Phi}_{\rm bulge},\,{\Phi}_{\rm thin},\,{\Phi}_{\rm thick}$ and ${\Phi}_{\rm halo}$,

\begin{equation}
\Phi (R,z) = {\Phi}_{\rm bulge}(r) + {\Phi}_{\rm thin}(R,z) + {\Phi}_{\rm thick}(R,z) + {\Phi}_{\rm halo}(r),
\end{equation}

and velocity contributions to the RC from different components are given by,

\begin{equation}
V^2_{\rm c} (R) = V^2_{\rm bulge} (R) + V^2_{\rm thin} (R) + V^2_{\rm thick} (R) + V^2_{\rm halo} (R).
\end{equation}

The Navarro-Frenk-White (NFW)  model (Navarro et al. 1996, 1997) which is derived from the simulations in the
CMD scenario of galaxy formation has been widely used for modeling the dark matter halo
(e.g., Sofue 2012; Wang et al. 2018).
We assume that density distributions of all stellar components are well-known, and the velocity contribution of the dark matter halo is
fitted by searching for the best-parameters by using the Markov Chain Monte Carlo method.
For the fitting model, the Miyamoto-Nagai potential model (Miyamoto \& Nagai 1975) and a spherical Plummer potential
(Plummer 1911) are used for the thin/thick disks and bulge, respectively. We take the parameter values
of the enclosed mass, the scale length, and the scale height from the model I of Pouliasis et al.(2017).

The NFW dark matter density profile is described as,

\begin{equation}
\rho(r) = \frac{\rho_{\rm crit}\,\delta_c}{(r/r_{\rm s})(1+r/r_{\rm s})^2},
\end{equation}
where $\rho_{\rm crit} = {3H^2}/8\pi G$, and $H=70.6\,{\rm km\,s^{-1}\,Mpc^{-1}}$ is taken for the Hubble constant.
The quantity of $\delta_c$ is the characteristic overdensity of the halo. Here, $r_{\rm s} = R_{\rm vir}/c$ is the scale radius, where $c$
is so-called concentration parameter, and $R_{\rm vir}$ is the virial radius. $R_{\rm vir}$ is related to the virial mass
as $M_{\rm vir}=200\rho_{\rm crit}\frac{4\pi}{3} {R^3_{\rm vir}} $ (see Navarro et al. (1996, 1997) for more details).
In the next subsections, the rotation curves from different kinematical
models and fitting results are discussed.


\subsection{The rotation curve from proper motions}

After measuring the proper motion of the star and setting the solar rotation speed as $V_{\rm c,\sun}=233.6\pm2.8\,{\rm km\,s^{-1}}$ (Mr$\acute{\rm o}$z et al. 2019),
then assuming a circular orbit for the Cepheid, the following formula gives the rotation velocity (Gnaci$\acute{\rm n}$ski 2019),

\begin{equation}
V_{\rm c} =   \frac{R}{{{\rm R}_\odot}{\rm cos}l - D} (V_{\rm t} + V_{\rm c,\sun} {\rm cos}l),
\end{equation}
where the transverse velocity $V_{\rm t} = D\mu_l$, and $\mu_l$ is the proper motion in the Galactic longitude (multiplied by ${\rm cos}b$).
The stars with $|z|>0.5$ kpc are excluded,
and unphysical velocities caused by small or negative denominators are removed
(see Gnaci$\acute{\rm n}$ski (2019) for the same selection criterion), so only 591
classical Cepheids are left from whole classical Cepheids for this kinematical modeling. Among our sample, there
are 168, 324 and 411 classical Cepheids distributed in the Galactocentric range of
$R>12$ kpc, $R>10$ kpc and $R>8$ kpc, respectively.
Figure 2 shows $\mu_l$ and the calculated rotation velocities of 591 classical Cepheids.
More than 98\% of $\mu_l$ have uncertainties less than 0.2 $\rm mas\,yr^{-1}$, and this leads to small uncertainties in the rotation velocity calculation.
The number of analyzed classical Cepheids in this work is about twice that used in
Gnaci$\acute{\rm n}$ski (2019), and we have more stars in the outer disc which is helpful to improve the accuracy of the RC measurement.

Figure 3 shows the rotation velocity distribution from $R = 4$ kpc and 19 kpc (see Table 1 for the values), and the linear function fitted from it is,

\begin{equation}
V_{\rm c} (R) = (222.91\pm2.08)\, {\rm km\,s^{-1}} + (-1.45\pm0.16)\,{\rm km\,s^{-1}\,kpc^{-1}}\times (R-{\rm R}_\odot).
\end{equation}

This yields a gently declining rotation curve with a small gradient of ($-1.45\pm0.16$) ${\rm km\,s^{-1}\,kpc^{-1}}$, and indicates the rotation
velocity at the position of the Sun as $V_{\rm c} ({R}_\odot) = 222.91\pm2.08\, {\rm km\,s^{-1}}$.
By fixing the contributions of baryonic components of the Galaxy (see model I of Pouliasis et al.(2017)), we estimated the mass and the
properties of the Milky Way's dark matter halo with the NFW profile (fitted results in Figure 3), and we derived
$M_{\rm vir} = (6.63\pm0.67)\times 10^{11}\,M_\odot$, corresponding to a viral radius $R_{\rm vir}=178.57\pm5.42$ kpc. We obtained the concentration of $c=12.36\pm0.42$
and a scale radius of $r_{\rm s}=14.45\pm0.46$ kpc. The indicated characteristic density is $\rho_0 = (1.05\pm0.12)\times10^{7}$ ${M_\odot\, \rm kpc^{-3}}$,
and dark matter density at the location of the Sun is $\rho_{\rm DM,\odot} = 0.28\pm0.04$ GeV ${\rm cm^{-3}}$.

\subsection{The rotation curve from 3D velocity vector}

The rotation velocity can also be determined from the 3D velocity vector if the three quantities of radial
velocity and proper motions are available. Reid et al. (2009) described
the calculation formulas of stellar motions by using radial velocity ($V_{\rm h}$) and proper motions, which we adopt here:
$U$--velocity component toward the Galactic center, $V$--velocity component along with the Galactic rotation, $W$--toward
the North Galactic pole. The optimizing model of $V_{\rm c} ({\rm R}) = V_{\rm c, \odot} + \frac{dV_{\rm c}}{dR}(R-\rm R_\odot)$, where $V_{\rm c,\odot}$ and
$\frac{dV_{\rm c}}{dR}$ are the Sun's rotation velocity and fitted parameter, is adopted for deriving rotation velocities
(see Reid et al. (2009) for more details).
For the peculiar (noncircular) solar motions with respect to the
local standard of rest, the values of $U_\odot = 11.1\pm1.3$ ${\rm km\,s^{-1}}$, $V_\odot = 12.24\pm2.1$ ${\rm km\,s^{-1}}$ and $W_\odot = 7.3\pm0.7$ ${\rm km\,s^{-1}}$
are taken from Sch$\ddot{\rm o}$nrich et al. (2010).

The proper motions of the sample are obtained from the Gaia DR2, and the radial velocities are derived by
cross-matching with Gaia DR2 and LAMOST DR6 data (e.g., Zhao et al. 2006, 2012). We excluded five
Cepheids known in the binary systems, and we put extra
constraints of $|z|\le2.0$ kpc and radial velocity uncertainty $<20$ ${\rm km\,s^{-1}}$ to remove 11 objects in order to reduce uncertainties.
It is well known that the radial velocity uncertainty may be larger for a single star when it's
measured near the pulsation phase (Stibbs 1955). However, the uncertainties of variable stars caused by the pulsation need further investigations, and
it may not clearly affect the statistical
result (see Ablimit \& Zhao 2017). For the 3D velocity model, we have 1078 classical
Cepheids: 836 of them from Skowron et al. (2019a,b), 55 from ASAS-SN catalog, 73 from Gaia DR2 Cepheids catalog, 22 from Chen et al.(2019),
and 92 of them are from Chen et al.(2020). Among our sample, there are 47, 165, 377 and 659 classical Cepheids distributed in the Galactocentric range of $R>14$ kpc,
$R>12$ kpc, $R>10$ kpc and $R>8$ kpc, respectively. In this work, the farest distance up to $\sim$19 kpc,
simply because no star satisfies the criterion to model beyond 19 kpc.
The radial velocities of 1043 stars are derived from the Gaia DR2 catalog while 35 of them obtained from LAMOST DR6.

$ Cleaned\, Sample.$ There are likely some objects in 1078 star sample, which may be members of binary systems (and unrecognized with incorrect
astrometric solutions) or categorized erroneously as classical Cepheids which as such and may actually be another type of variable.
There are also some classical Cepheids with observed velocity components about 4$\sigma$ ($\sigma$ is dispersion of residuals)
larger than the mean. Considering these possibilities and uncertainties,
we selected 963 classical Cepheids from 1078 stars as the cleaned sample,
and derived rotation velocities of the cleaned sample are shown in Table 1.
The measured RCs from the cleaned sample and all 1078 sample can be fitted by the same linear function.

The distributions of $V_{\rm h}$, $\mu_l$, $\mu_b$ and rotation velocities are given in Figure 4.
The rotation curve from the 3D velocity vector (Figure 5) is well approximated by the following linear function,
\begin{equation}
V_{\rm c} (R) = (232.5\pm0.83)\, {\rm km\,s^{-1}} + (-1.33\pm0.1)\,{\rm km\,s^{-1}\,kpc^{-1}}\times (R-{\rm R}_\odot).
\end{equation}

The rotation curve from this method is gently decreasing with a derivative of ($-1.33\pm0.1$) ${\rm km\,s^{-1}\,kpc^{-1}}$. The slope of the curve and rotation
velocity at the location of the Sun ($V_{\rm c} ({R}_\odot) = 232.5\pm0.83\, {\rm km\,s^{-1}}$) are in a good agreement
with the results of Mr$\acute{\rm o}$z et al. (2019), as about 70\%  of data points in the sample overlap with that of Mr$\acute{\rm o}$z et al.
(2019). However, there are more than 300 different stars in this work. In particular, our sample has more stars in the outer disc, which improves the accuracy of the RC,
and is helpful to put more accurate constraint on the distribution of dark matter in the Milky Way.
Comparing to the virial mass from the proper motion method, we derived a higher viral mass in this method,
$M_{\rm vir} = (8.22\pm0.52)\times 10^{11}\,M_\odot$ with a corresponding viral radius $R_{\rm vir}=191.84\pm4.12$ kpc.
The resulted concentration and scale radius are $c=13.04\pm0.34$ and $r_{\rm s}=14.71\pm0.42$ kpc, respectively.
The estimated characteristic density and dark matter density at the location of the Sun
are $\rho_0 = (1.20\pm0.1)\times10^{7}$ ${M_\odot\, \rm kpc^{-3}}$ and
$\rho_{\rm DM,\odot} = 0.33\pm0.03$ GeV ${\rm cm^{-3}}$, respectively.

 \subsection{Comparison and discussion}

 There are 366 common classical Cepheids modeled in two methods, and different tracers are
 selected due to different criterion for different methods. The discrepancy of two methods' results are within 10\%.
 The most important advantage of our sample is the accuracy of the distances which have uncertainties at a level of 2-3\%.
 We have small uncertainties in our results (see values of uncertainties in Table 1), however, only bootstrapping uncertainties without the systematic uncertainties are considered
in this work (see Eilers et al. (2019) for analysis of the possible systematic uncertainties). The effect of the asymmetric drift
is not considered in the calculation of this work due to the very small systematic uncertainty it causes (e.g., estimated as $\pm$0.28 $\rm km\,s^{-1}$ by Kawata et al. 2018). Within 19 kpc,
all systematic uncertainties added up (i.e. caused by uncertainties of distances, uncertainty of ${R}_\odot$ and the asymmetric drift etc.)
only affect the RC measurement at a $\lesssim 5\%$ level. It is well known that the motions of stars are affected by Galactic
substructures (e.g. Grand, Kawata \& Cropper 2014; Bovy et al.2015; Kawata et al. 2018;  Martinez-Medina et al. 2019).
We did not use stars located at $R < 4.0$ kpc in order to reduce the influence of other structures like the Galactic bar.

The slopes of the rotation curves from two methods are gently decreasing,
as favored by the recent discoveries (e.g., Mr$\acute{\rm o}$z
et al. 2019; Eilers et al. 2019). They are not as flat as demonstrated
in Sofue et al. (2009) and Reid et al. (2014), and it is not
as steep as showed in Gnaci$\acute{\rm n}$ski (2019).
This indicates that the dark matter content would not possibly so
high or so low claimed in those previous works.
The result (see the cross-point between the rotation curve of all stellar components
and dark matter halo in Figure 3) from the proper motion method suggests that the dark matter halo dominates
the Galactic rotation when $R\gtrsim 14.5$ kpc, and this is in good agreement with recent finding by Eilers et al. (2019).
However, based on the 3D velocity method (as shown in Figure 5), the dark matter halo dominates the rotation velocity if $R\gtrsim 12.5$ kpc.
The comparison of two velocity distributions from two methods give a
same dip-like feature, there are a clear declining at the distance around $\sim$10 kpc, and this is consistent with the similar
dip claimed by previous works (Sofue et al 2009; Kafle et al. 2012; McGauph 2018).
However, there is no dip in the results of the cleaned sample with the 3D velocity method (Eilers et al. 2019).

The rotation velocity of the Sun found in the proper motion method is
in good agreement with the results of some previous works (e.g., Bovy et al. 2012; Wegg et al. 2018).
The Sun's rotation velocity obtained from the 3D velocity method is, within uncertainties, consistent
with the relatively higher values reported by Metezger et al. (1998), Reid et al.(2014), Kawata et al.(2018) and Mr$\acute{\rm o}$z et al. (2019).
The estimated virial masses from the two methods in this work are lower than the values ($(\sim 1.0-\sim 2.0)\times 10^{12}\,M_\odot$) derived by K$\ddot{\rm u}$pper et al. (2015),
Bland-Hawthorn \& Gerhard (2016), Watkins et al. (2019), Callingham et al.(2019) and Li et al. (2019).
Within uncertainties, the viral masses in this work are in good agreement with the results of Bovy et al. (2012),
Kalfe et al. (2012), Eadie et al. (2018), Eadie \& Juri$\acute{\rm c}$ (2019), Eilers et al.(2019) and Cautun et al.(2019).
The estimated mass by our 3D velocity method has very good agreement with the mean viral mass ($({0.83_{-0.09}^{+0.12}})\times 10^{12}\,M_\odot$)
derived by Karukes et al. (2019).

The estimated local dark matter densities from two methods including the uncertainties
are consistent with the values of Weber \& de Boer, (2010), Sofue (2012), Eilers et al. (2019) and Callingham et al.(2020). However,
they are higher than the values ($<\sim0.2$ GeV ${\rm cm^{-3}}$) given by Gnaci$\acute{\rm n}$ski (2019), while they are lower than the
estimated density ($\sim 0.9$ GeV ${\rm cm^{-3}}$) by Garbari et al.(2012) and ($0.542\pm0.042$ GeV ${\rm cm^{-3}}$) by Bienaym$\acute{\rm e}$ et al. (2014).

Effect of uncertainties of baryonic mass components. The estimation of the dark matter
halo profiles rely on observational results of the baryonic mass components. Recently, de Salas et al.(2019) discussed that the dark matter density
estimation is more sensitive to the uncertainties of the baryonic components rather than the uncertainties of
the rotation velocities. They found a different uncertainty ($\pm0.149$ GeV ${\rm cm^{-3}}$) of the dark matter density with the same velocities, and it is
about 3 times of what Eilers et al. (2019) find. They also show that using different model such as the NFW dark matter profile and  Einasto
dark matter profile also gives a uncertainty of $\pm0.036$ GeV ${\rm cm^{-3}}$.
Comparing to the Galactic disc mass in some previous works (e.g., Smith et al. 2007),  we took relatively higher masse for the Galactic (thin $+$ thick) disc
from Pouliasis et al.(2017). Thus, we may underestimate the halo profiles. We examine it by taking a very simple example, and we run the model by
using $5.0\times 10^{10}\,M_\odot$ for the whole disc mass instead of
$7.888\times 10^{10}\,M_\odot$ (thin $+$ thick discs) as in this work from Model I of Pouliasis et al.(2017). We found that the dark matter density goes up to
$0.408$ GeV ${\rm cm^{-3}}$ when we reduce the baryonic mass of Galactic disc in the estimation modeling, and it is $0.078$ GeV ${\rm cm^{-3}}$ higher than the
value ($0.33$ GeV ${\rm cm^{-3}}$) derived from Model I of Pouliasis et al.(2017).
This supports the statement given by de Salas et al.(2019).
The future observational data may provide better constraints on the baryonic components\footnote{Note that previous works made important
progress in modeling the baryon budget of Galactic disc and its uncertainties (e.g., Flynn et al. 2006; Bovy \& Rix 2013)}.

Interestingly, the density derived
from our 3D velocity method is basically consistent with the estimated dark matter density by de Salas et al.(2019) which is in
a range of $(0.3-0.4)$ GeV ${\rm cm^{-3}}$. Our local dark matter density estimated from 3D velocity method is in very good agreement with the
local dark matter density ($0.32-0.36$ GeV ${\rm cm^{-3}}$) inferred from fitting models to the Gaia DR2 Galactic rotation
curve and other data (Cautun et al. 2019).

\section{Conclusion}

We have analyzed 3483 classical Cepheids selected from thousands of classical Cepheids identified by several survey projects (e.g., OGLE, ASAS-SN, Gaia, WISE and ZTF),
and constructed the rotation velocity distribution of the Milky Way between the Galactocentric distance 4 kpc and 19 kpc by
using two different methods. The distances of these classical Cepheids have the typical uncertainties of $< 3\%$ (which is crucial in the analysis of the rotation curve), and 3D
spatial distributions show a vary clear Galactic warp feature claimed by previous works (see the section 2).
591 and 1078 classical Cepheids have been analyzed by using the proper motion and 3D velocity methods,
and most of observed uncertainties of proper motions and radial velocities are less than 0.2 $\rm mas\,yr^{-1}$ and 20 $\rm km\,s^{-1}$, respectively.
This represents the largest classical Cepheid sample analyzed to date. We apply the NFW profile approach to simulate
the dark matter content of the Milky Way. Our main findings are,

\begin{enumerate}

\item{The different methods or/and different sample would give different results in some extent.
The uncertainties of baryonic components also have important role in the estimation of dark matter profiles.
The result of the proper motion method shows that the dark matter halo is main contributor to the
Galactic rotation when the distance $R\gtrsim 14.5$ kpc,
while the 3D velocity modeling demonstrates that the Galactic rotation curve is dominated
by the dark matter halo at $R\gtrsim 12.5$ kpc.
The rotation curve constructed by both method are gently declining. The rotation curve
from 3D velocity method is decreasing more gently with a derivative of
($-1.33\pm0.1$) ${\rm km\,s^{-1}\,kpc^{-1}}$.
The rotation velocity at the position of the Sun (($232.5\pm0.83$) ${\rm km\,s^{-1}}$) obtained from the 3D velocity method
is about $10\,\rm km\,s^{-1}$ faster than the rotation velocity of the Sun derived from the proper motion method.}

\item{The best-estimation with the NFW profile based on the rotation curve of the 3D velocity
method generates a higher viral mass ($M_{\rm vir} = (0.822\pm0.052)\times 10^{12}\,M_\odot$)
with the corresponding radius of $R_{\rm vir}=191.84\pm4.12$ kpc and concentration of $c=13.04\pm0.34$.
At the same time, the predicted local dark matter
density ($\rho_{\rm DM,\odot} = 0.33\pm0.03$ GeV ${\rm cm^{-3}}$) is also higher than the estimated
value from the proper motion modeling.}

\end{enumerate}

\begin{acknowledgements}

We are grateful to Anna-Christina Eilers for useful discussions
and comments to improve the manuscript.
We thank Xiaodian Chen for providing the new classical Cepheids catalog identified
from the ZTF data.
This work was supported by National Natural Science Foundation of China under grant
number 11988101, 11890694 and the National Key R\&D Program of China No. 2019YFA0405502.
The LAMOST FELLOWSHIP is supported by Special Funding for Advanced Users, budgeted and
administrated by Center for Astronomical Mega-Science, Chinese Academy of Sciences.

This publication made use of data from the European Space Agency (ESA) mission Gaia (https://www.cosmos.esa.int/ gaia),
processed by the Gaia Data Processing and Analysis Consortium (DPAC, https://www.cosmos.esa.int/web/gaia/ dpac/consortium).
Funding for the DPAC has been provided by national institutions, in particular the institutions participating in the Gaia Multilateral Agreement.
This work has used the data products from the Wide field Infrared Survey Explorer (WISE), which is a joint project of the University of California,
Los Angeles, and the Jet Propulsion Laboratory/California Institute of Technology, funded by the National Aeronautics and Space Administration.
The work also have used the data from the Large Sky Area Multi-Object Fiber Spectroscopic Telescope (LAMOST) which is a National Major
Scientific Project built by the Chinese Academy of Sciences. Funding for the project has been provided by the National Development and
Reform Commission. LAMOST is operated and managed by the National Astronomical Observatories, Chinese Academy of Sciences.

\end{acknowledgements}

\begin{center}
 REFERENCES
\end{center}

Ablimit, I., \& Zhao, G. 2017, ApJ, 846, 10

Ablimit, I., \& Zhao, G. 2018, ApJ, 855, 126

Bellm, E. C., Kulkarni, S. R., Graham, M. J., et al. 2019, PASP, 131, 018002

Bhattacharjee, P., Chaudhury, S., \& Kundu, S. 2014, ApJ, 785, 63

Bienaym$\acute{\rm e}$, O., Famaey, B., Siebert, A., et al. 2014, A\&A, 571, A92

Binney, J., \& Wong, L. K. 2017, MNRAS, 467, 2446

Bland-Hawthorn, J., \& Gerhard, O. 2016, ARA\&A, 54, 529

Bosma, A., \& van der Kruit ,P. C. 1979, A\&A, 79, 281

Bovy, J., Allende Prieto, C., Beers, T. C. et al., 2012, ApJ, 759, 131

Bovy, J., \& Rix, H.-W., 2013, ApJ, 779, 115

Bovy J., 2015, ApJS, 216, 29

Bowden, A., Belokurov, V., Evans, N. W. 2015, MNRAS, 449, 1391

Callingham, T. M., Cautun, M., Deason, A. J., et al. 2019, MNRAS, 484, 5453

Callingham, T. M., Cautun, M., Deason, A. J., et al. 2020, eprint arXiv:2001.07742

Cautun, M., Benitez-Llambay, A., Deason, A. J., et al. 2019, eprint arXiv:1911.04557

Chen, X. D., Wang, S., Deng, L. C., et al. 2019,
Nature Astronomy, 3, 320

Chen, X. D., Wang, S., Deng, L. C., et al. 2020, submitted to AAS Journals

Deason,A. J., Belokurov, V. \& Sanders, J. L. 2019, arXiv:1912.02599v2

Dubinski, J. 1994, ApJ, 431, 617

Eadie, G., \& Juri$\acute{\rm c}$, M. 2019, ApJ, 875, 159

Eadie, G. M., Keller, B. \& Harris, W. E. 2018, ApJ, 865, 72

Eilers, A.-C., Hogg, D. W., Rix, H.-W., Ness, M. 2019, ApJ, 871, 120

Flynn, C., Holmberg, J., Portinari, L. et al. 2006, MNRAS, 372, 1149

Freeman, K. C. 1970, ApJ, 160, 811

Frink, S., Fuchs, B. \& Wielen, R. 1995, Astronomische Gesellschaft Abstract Series 11, 196

Gaia Collaboration, Prusti, T., de Bruijne, J. H. J., et al. 2016, A\&A, 595, A1

Gaia Collaboration, Abuter, R., Amorim, A., et al. 2018, A\&A, 615, L15

Garbari, S., Liu, C., Read, J. I. \& Lake, G. 2012, MNRAS, 425, 1445

Gnaci$\acute{\rm n}$ski, P., 2019, AN, 340, 787

Grand R. J. J., Kawata D., Cropper M., 2014, MNRAS, 439, 623

Gravity Collaboration, Abuter, R., Amorim, A., et al. 2018, A\&A, 615, L15

Gunn, J. E., Knapp, G. R., \& Tremaine, S. D. 1979, AJ, 84, 1181

Honma, M., Bushimat, T., Choi, Y. K. et al. 2007, PASJ, 59, 889

Ibata, R., Lewis, G. F., Irwin, M., Totten, E., Quinn, T. 2001, ApJ, 551, 294

Jayasinghe, T., Stanek, K. Z., Kochanek, C. S. et al. 2018, arXiv:1809.07329

Kafle, P. R., Sharma, S., Lewis, G. F. \& Bland-Hawthorn, J. 2012, ApJ, 761, 98

Kafle, P. R., Sharma, S., Lewis, G. F. \& Bland-Hawthorn, J. 2014, ApJ, 794, 59

Karukes, E. V.,  Benito, M., Iocco, F., et al. 2019, arXiv:1912.04296v1

Katz, D., Antoja, T., et al. 2018 A\&A, 616, A11

Kawata, D., Bovy, J., Matsunaga, N., \& Baba, J. 2018, MNRAS, 482, 40

K$\ddot{\rm u}$pper, A. H. W., Balbinot, E., Bonaca, A., et al. 2015, ApJ, 803, 80

Lake, G. 1989, AJ, 98, 1554

Levine, E. S., Heiles, C., \& Blitz, L. 2008, ApJ, 679, 1288

Li, Z. Z., Qian, Y. Z., Han, J., et al. 2019, arXiv:1912.02086v1

L$\acute{o}$pez-Corredoira, M., Abedi, H., Garz$\acute{o}$n, F. \& Figueras, F. 2014, A\&A, 572, 101

Lux, H., Read, J. I., Lake, G., Johnston, K. V. 2012, MNRAS, 424, L16

Martinez-Medina, L.,  Pichardo, B.,  Peimbert, A. \&  Valenzuela, O. 2019, MNRAS, 485, L105

McGauph, S. S. 2018, Reseach Notes of the American Astronomical Society, 2, 156

Medina, G. E., Munoz, R. R., Vivas, A. K., et al. 2018. ApJ, 855, 43

Mel'nik, A. M., Rautiainen, P., Berdnikov, L. N., Dambis, A. K., Rastorguev, A. S. 2015, AN, 336, 70

Metezger, M. R.,  Caldwell, J. A. R. \&   Schechter, P. L. 1998, ApJ, 115, 635

Miyamoto, M. \& Nagai, R. 1975, PASJ, 27, 533

Monari, G., Famaey, B., Carrillo, I. et al., 2018, A\&A, 616, L9

Mr$\acute{\rm o}$z, P., Udalski, A., Skowron, D. M., et al.
2019, ApJL, 870, L10

Navarro, J. F., Frenk, C. S., \& White, S. D. M. 1996, ApJ, 462, 563

Navarro, J. F., Frenk, C. S., \& White, S. D. M. 1997, ApJ, 490, 493

Nesti, F., \& Salucci, P. 2013, JCAP, 07, 016

Pato, M., \& Iocco, F. 2017, SoftX, 6, 54

Plummer, H. C. 1911, MNRAS, 71, 460

Pont, F., Queloz, D., Bratschi, P., \& Mayor, M. 1997, A\&A, 318, 416

Pouliasis, E., Di Matteo, P. \& Haywood, M. 2017, A\&A, 598, 66

Read, J. I., Lake, G., Agertz, O., Debattista, V. P. 2008, MNRAS, 389, 1041

Reid, M. J., Menten, K. M., Zheng, X. W., et al. 2009, ApJ, 700, 137

Reid, M. J., Menten, K. M., Brunthaler, A., et al. 2014, ApJ, 783, 130

Ripepi, V.£¬ Molinaro, R.£¬ Musella, I. et al. 2019, 2019A\&A, 625A, 14R

Russeil, D., Zavagno, A., Mege, P., et al. 2017, A\&A, 601, L5

Sch$\ddot{\rm o}$nrich, R.,  Binney, J., \& Dehnen, W. 2010, MNRAS, 403, 1829

Shappee, B. J., Prieto, J. L., Grupe, D. et al. 2014, ApJ, 788, 48

Skowron, D. M., Skowron, J., Mroz, P., et al. 2019a, Science, 365, 478

Skowron, D. M., Skowron, J., Mroz, P., et al. 2019b, Acta Astronomica, 69, 305

Smith, M. C., Ruchti, G. R., Helmi, A., et al. 2007, MNRAS, 379, 755

Sofue, Y., Honma, M., \& Omodaka, T. 2009, PASJ, 61, 227

Sofue, Y. 2012, PASJ, 64, 75

Sohn, S. T., Watkins, L. L., Fardal, M. A., et al. 2018, ApJ, 862, 52

Stibbs, D. W. N. 1955, MNRAS, 115, 363

Udalski, A., Szymanski, M. K. \& Szymanski, G. 2015 Acta Astron. 65, 1

Udalski, A., Soszynski, I., Pietrukowicz, P., et al. 2018, Acta Astronomica, 68, 315

Utkin, N. D., Dambis, A. K., Rastorguev, A. S., Klinchev, A. D., Ablimit, I., \& Zhao, G. 2018, Astronomy Letter, 44, 688

van Albada, T. S., Bahcall, J. N., Begeman, K., Sancisi, R. 1985, ApJ, 295, 305

Volders, L. M. J. S. 1959, Bull. Astron. Inst. Netherlands, 14, 323

Wang, W. T., Han, J. X., Cole, S., et al. 2018, MNRAS, 476, 5669

Watkins, L. L., van der Marel, R. P., Sohn, S. T. \& Evans, N. W. 2019, ApJ, 873, 118

Weber, M., \& de Boer, W. 2010, A\&A, 509, A25

Wilkinson, M. I. \& Evans, N. W. 1999, MNRAS, 310, 645

Wright, E. L., Eisenhardt, P. R. M., Mainzer, A. K., et al. 2010, AJ, 140, 1868

Zhao, G., Chen, Y.-Q., Shi, J.-R., et al. 2006, ChJAA, 6, 265

Zhao, G., Zhao, Y. H., Chu, Y. Q., et al. 2012, RAA, 12, 723


\begin{table}

\begin{center}
\caption{Measurements of the Galactic rotation velocity based on two different methods. For the 3D velocity method, the results of cleaned sample are given in the table.}

\begin{tabular}{cccccc}
 \hline\hline
 & Proper motion method &  & & 3D velocity method \\

 \hline
R (kpc) &$V_{\rm C} ({\rm km\,s^{-1}})$ &$\sigma_{V_{\rm C}} ({\rm km\,s^{-1}})$ |&R (kpc) & $V_{\rm C} ({\rm km\,s^{-1}})$ &$\sigma_{V_{\rm C}} ({\rm km\,s^{-1}})$\\
\hline
 4.2 & 234.11 & 3.96&   4.56 & 230.15 & 7.15\\
 4.8 & 241.24 & 2.75&   5.32 & 234.93 & 8.01\\
5.6 & 246.45 & 2.61&  6.11 & 237.41 & 5.97\\
 6.5 & 244.43 & 3.49 &    6.97 & 236.21 & 4.67\\
 7.5 & 242.69 &  7.35  & 7.78 & 234.02 & 3.77\\
 8.5 & 213.65  &  1.91  & 8.59 & 232.51 & 2.68\\
9.5 & 206.04 & 2.03 &   9.33 & 231.42 & 2.17\\
10.5 & 207.26 & 2.21 &   10.11 & 231.61 & 1.99\\
11.5 & 213.31 & 2.39 &   10.88 & 229.08 & 1.95\\
12.5 & 210.75 & 2.37 &   11.67 & 226.93 & 2.04\\
13.5 & 211.49 & 2.38 &   12.36 & 226.61 & 1.55\\
14.5 & 214.88 & 2.49 &   13.04 & 225.63 & 2.11\\
15.5 & 219.08 & 2.58 &   13.86 &226.36 & 1.61\\
16.5 & 212.45 & 2.49 &   14.61 & 225.87 & 2.21\\
17.5 & 210.62 & 2.62 &   15.42 & 226.13 & 2.09\\
18.5 & 211.14  & 2.42 &   16.26 & 223.29 & 2.56\\
 &  &  &   17.04 & 219.46 & 0.30\\
 &  &  &   17.87 & 210.68 & 2.72\\
 &  &  &   18.62 & 216.15 & 4.76\\
\hline
\end{tabular}
\end{center}
\end{table}

\begin{figure}
\centering
\includegraphics[totalheight=2.4in,width=3.2in]{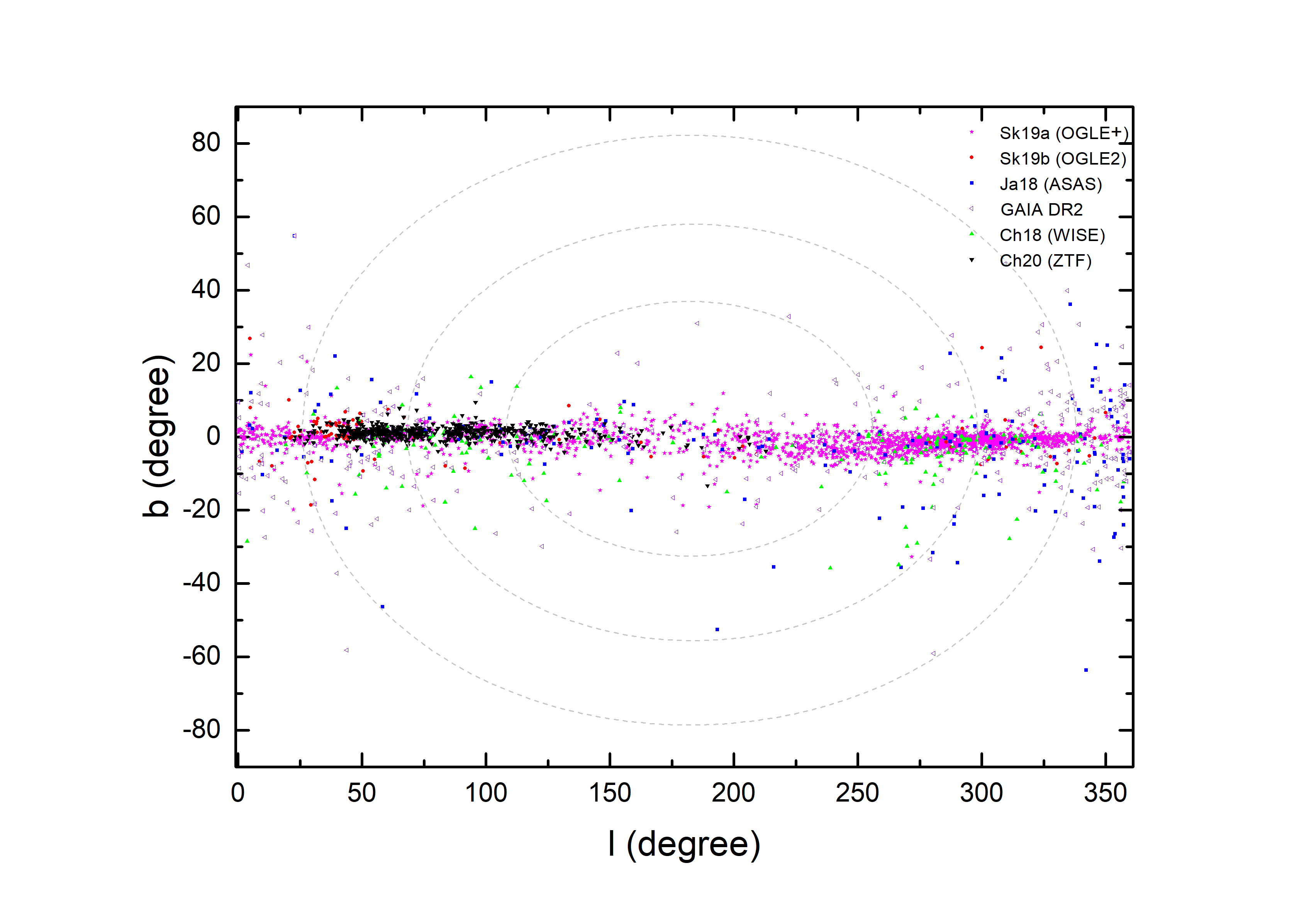}
\includegraphics[totalheight=2.4in,width=3.25in]{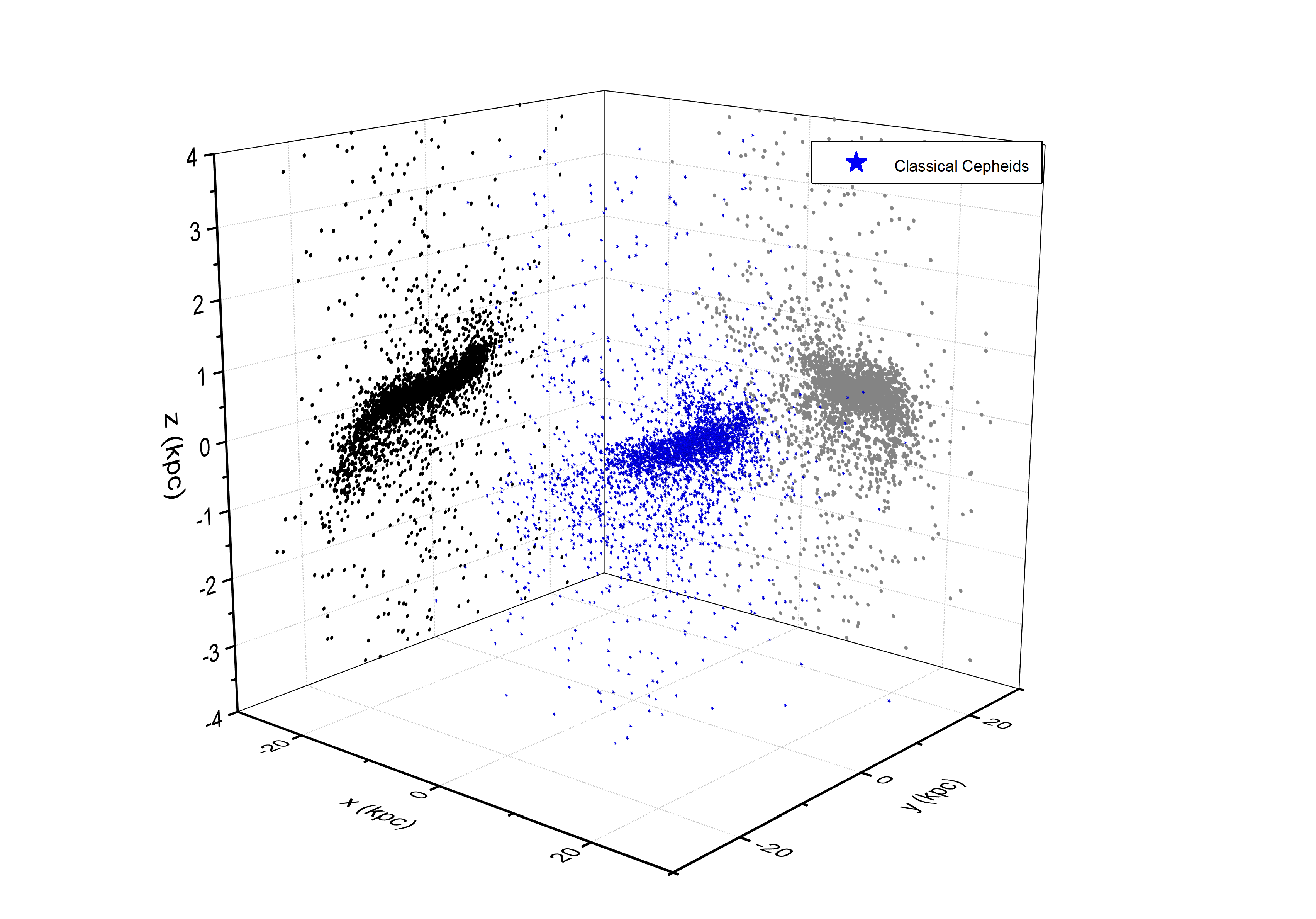}
\includegraphics[totalheight=2.4in,width=3.05in]{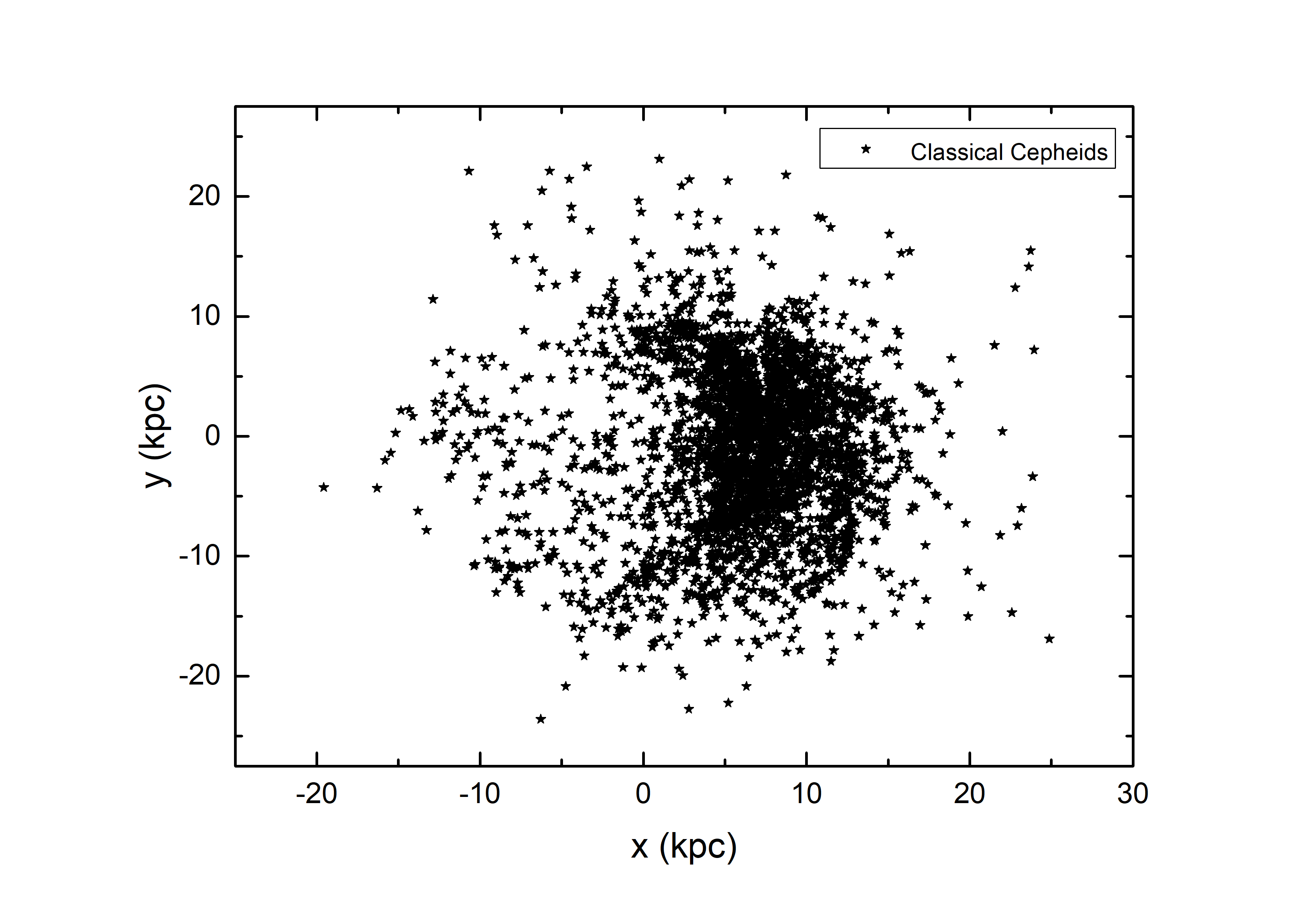}
\includegraphics[totalheight=2.4in,width=2.9in]{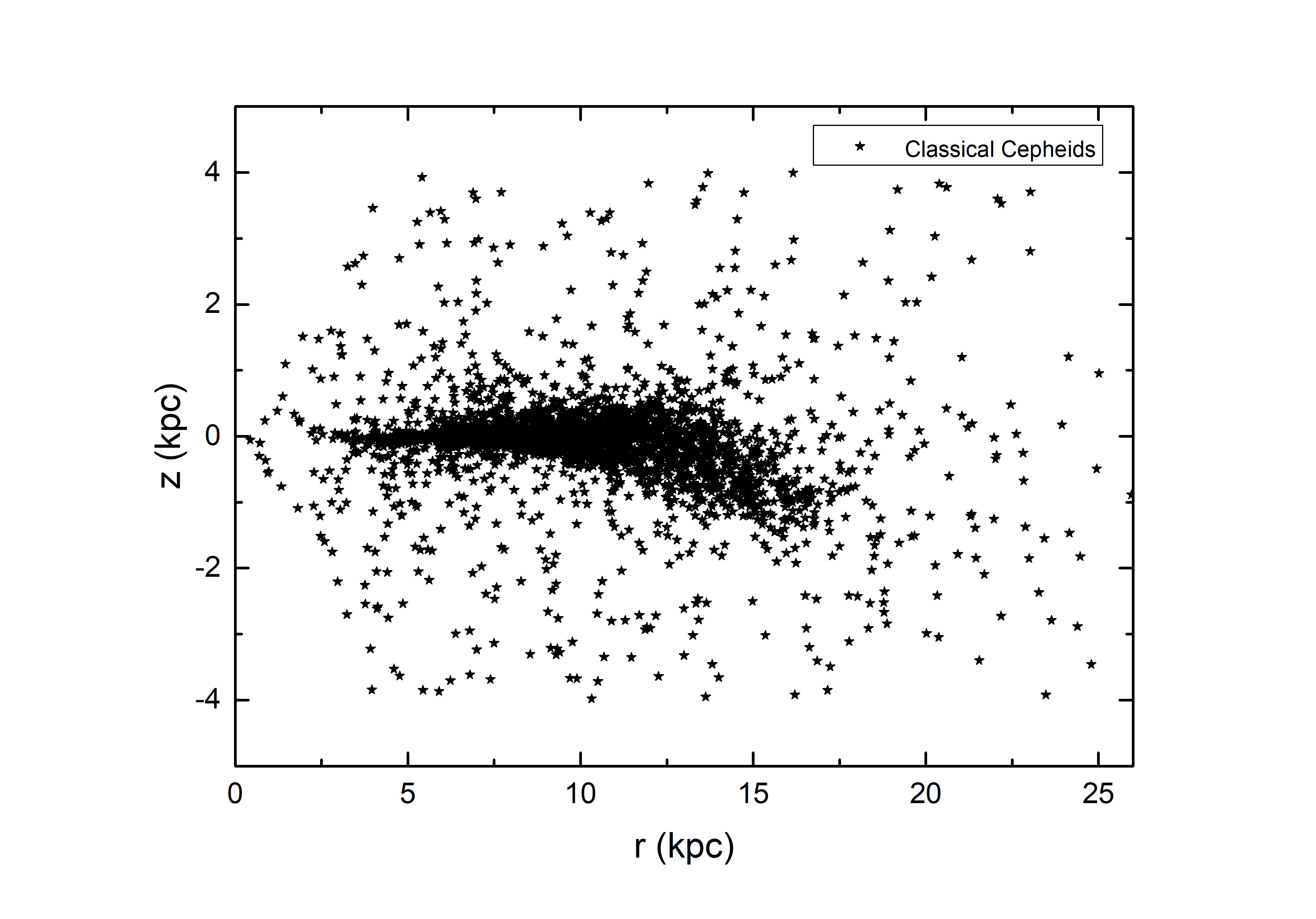}
\caption{The distributions of the Galactic longitude ($l$) \& latitude ($b$), and spatial distributions.
The 3D positions, projections in $x-z$ and $y-z$ planes are shown in the upper right panel; projection in $x-y$
and $r-z$ planes are given lower left and right panels, respectively.}
\label{fig:1}
\end{figure}

\clearpage

\begin{figure}
\centering
\includegraphics[totalheight=4.9in,width=6.2in]{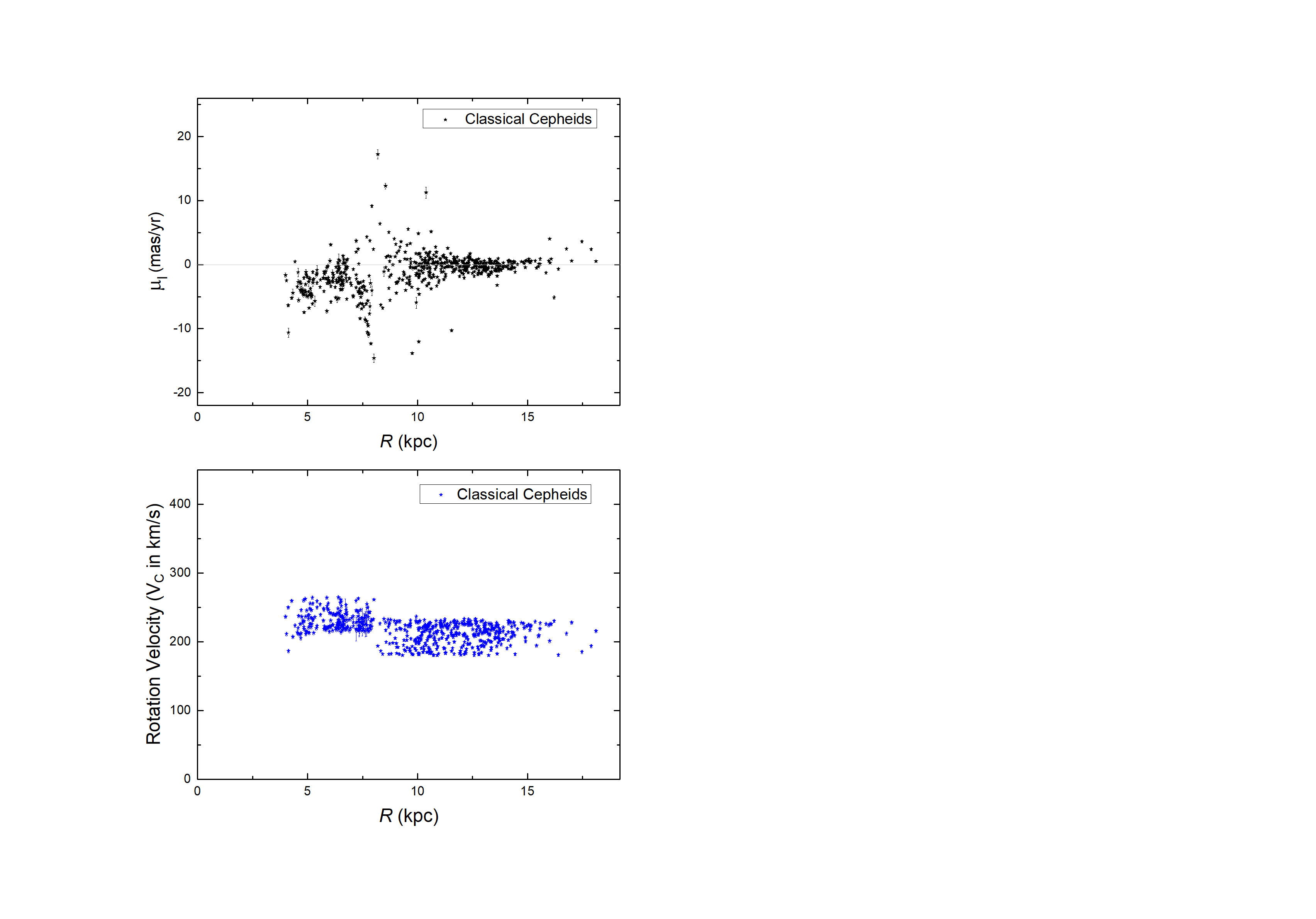}
\caption{The distributions of proper motions and derived rotation velocities in the proper motion model.}
\label{fig:1}
\end{figure}

\clearpage

\begin{figure}
\centering
\includegraphics[totalheight=4.5in,width=6.0in]{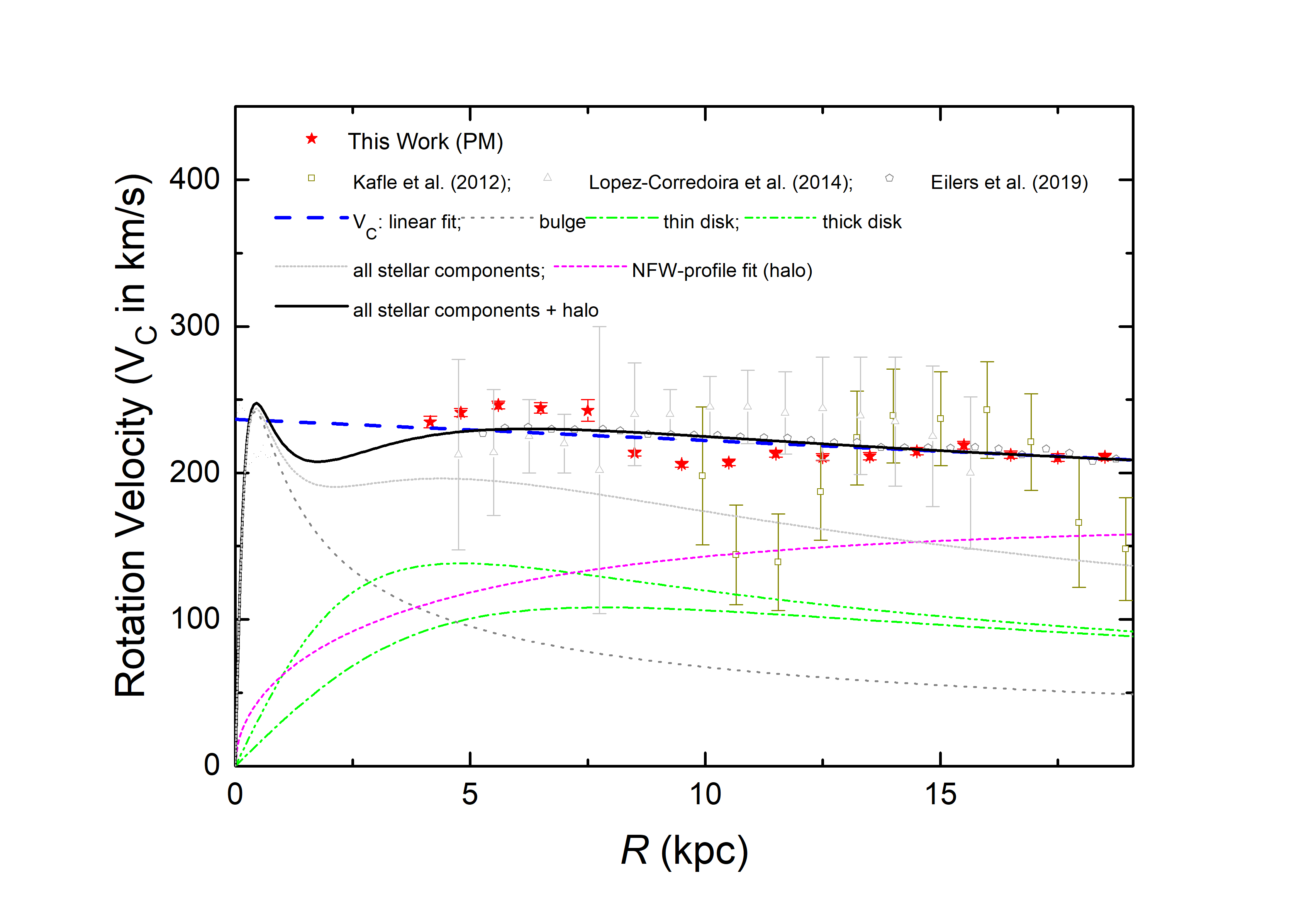}
\caption{The red stars show the distribution of new measured rotation velocities of the Milky Way from the proper motion method, and the error bar are derived
existing errors of the sample without including the systematic uncertainties. The blue dashed line is the linear fit to the new data in this work.
The black solid line is the best fit to the rotation velocity with a assumption that the Milky Way
components are the bulge (grey dotted line), thin disk  (green dash-dotted line), thick disk (green dash-dot-dotted line)
and dark matter halo (magenta short dashed line) by the NFW profile. The light grey short-dotted line represents the fit to
the rotation velocity modeled as sum of all stellar components.
The best-fit to the rotation velocity curve modeled as the sum of all components of the Milky Way is shown by the black solid line.
Three other symbols with different colors demonstrate the rotation velocities taken from three previous works for the comparison.}
\label{fig:1}
\end{figure}

\clearpage

\begin{figure}
\centering
\includegraphics[totalheight=6.0in,width=7.8in]{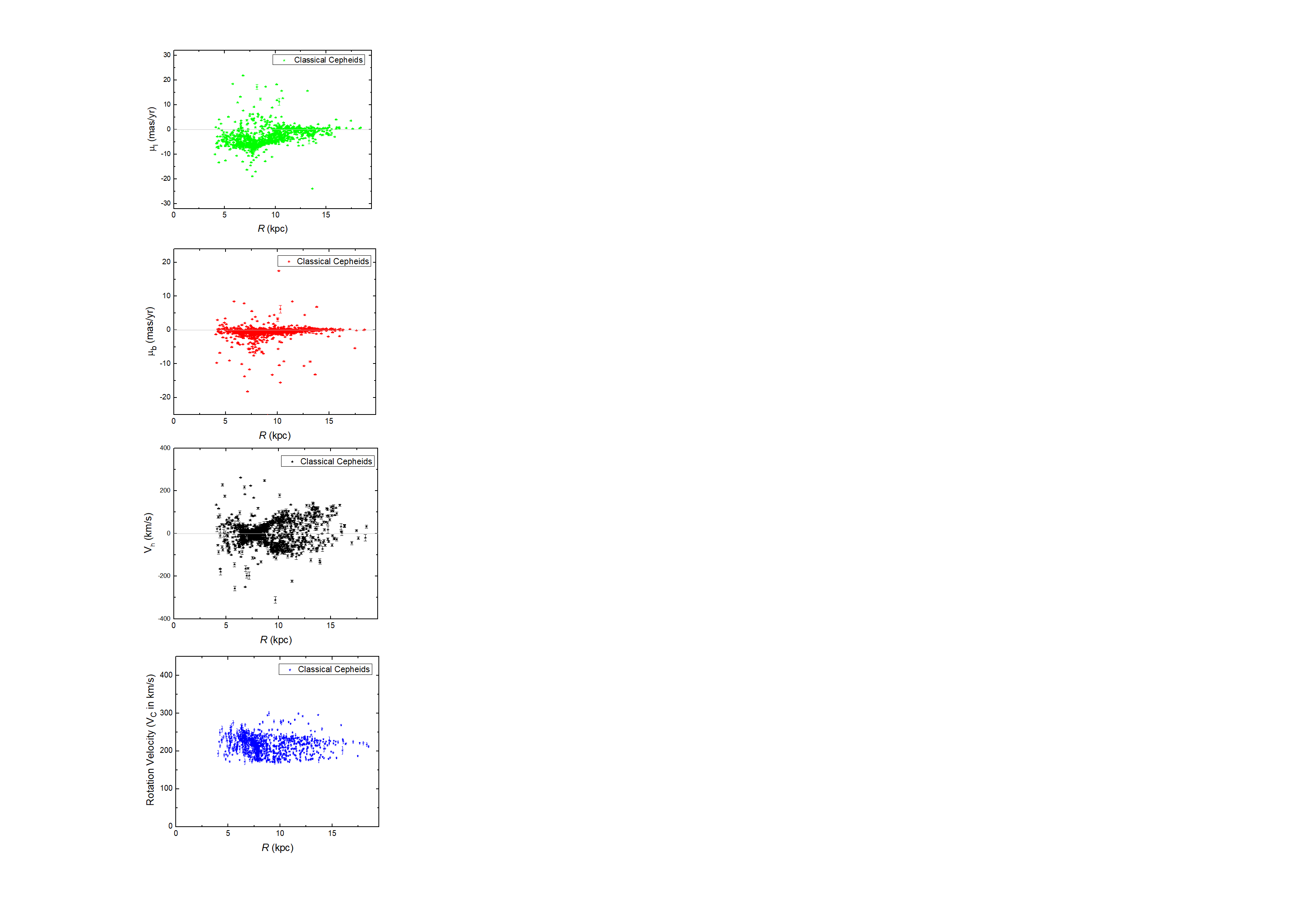}
\caption{The distributions of the proper motions in the Galactic longitude direction, proper motions
in the Galactic latitude direction and radial velocities used in the 3D velocity vector method.
The calculated rotation velocities of individual stars also shown in the figure.}
\label{fig:1}
\end{figure}

\clearpage

\begin{figure}
\centering
\includegraphics[totalheight=4.5in,width=6.0in]{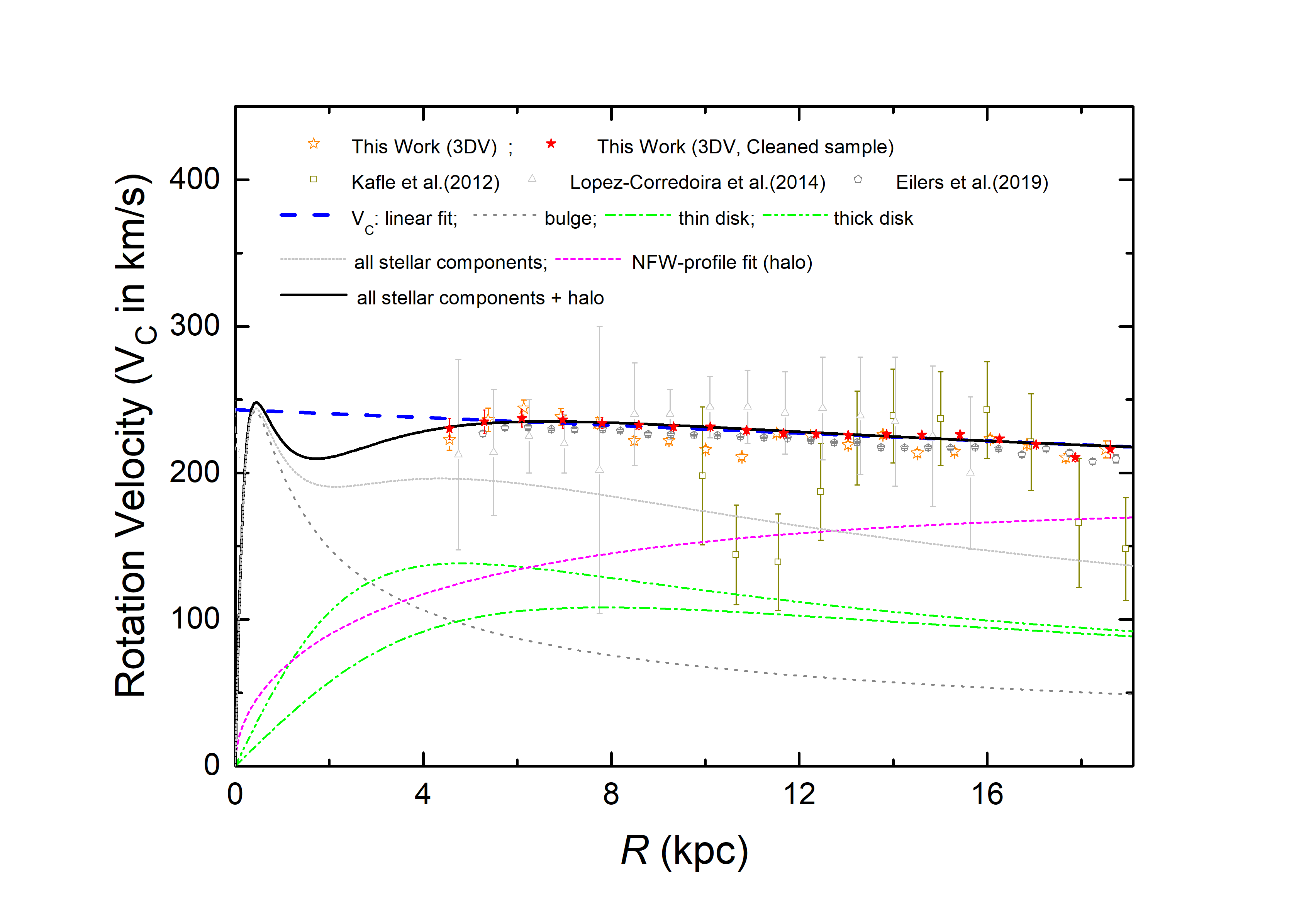}
\caption{The same as Figure 3, but based on the 3D velocity vector method, and the orange
open stars and red filled stars show the results of all 3D sample and cleaned sample, respectively.}
\label{fig:1}
\end{figure}

\clearpage

\end{document}